\documentclass[11pt]{emulateapj}
\usepackage{epsfig,apjfonts}

\def\beq{\begin{equation}}
\def\enq{\end{equation}}
\def\ba{\begin{eqnarray}}
\def\ea{\end{eqnarray}}
\def\Meszaros{M\'esz\'aros~}
\def\<{<\!\!}
\def\>{\!\!>}

\begin{document}

\title{Prompt TeV neutrinos from dissipative photospheres of $\gamma$-ray bursts}

\author{Xiang-Yu Wang and Zi-Gao Dai }

\affil{Department of Astronomy, Nanjing University, Nanjing
210093, China }

\begin{abstract}
Recently, it was suggested that a photospheric component that
results from the internal dissipation occurring in the optically
thick inner parts of relativistic outflows may be present in the
prompt $\gamma$/X-ray emission of gamma-ray bursts or X-ray flashes.
We explore high-energy neutrino emission in this dissipative
photosphere model, assuming that the composition of the outflow is
baryon-dominated. We find that  neutrino emission from proton-proton
collision process forms an interesting signature in the neutrino
spectra. Under favorable conditions for the shock dissipation site,
these low-energy neutrinos could be detected by ${\rm km^3}$
detectors, such as Icecube. Higher energies ($\ga10$ TeV) neutrino
emission from proton-proton collision and photo-pion production
processes could be significantly suppressed for dissipation at
relatively small radii, due to efficient Bethe-Heitler cooling of
protons and/or radiative cooling of the secondary mesons in the
photosphere radiation. As the dissipation shocks continue further
out, high energy neutrinos from photo-pion production process
becomes dominant.

\end{abstract}


\section {Introduction}
Although it has been generally accepted that the prompt gamma-ray
emission of gamma-ray bursts (GRBs) results from internal
dissipation, likely internal shocks, of a relativistic outflow
(e.g. Paczy{\'n}ski \& Xu 1994; Rees \& \Meszaros 1994), the
dissipation site and the radiation mechanism  for the gamma-ray
emission are still largely unknown. Synchrotron and/or
inverse-Compton scattering emission by shock-accelerated electrons
in the optically-thin region has been proposed as an efficient
mechanism for the gamma-ray emission. However, this model does not
satisfactorily account for a few observational facts, such as the
low-energy spectral slops that are steeper than synchrotron lower
energy spectral indices (Preece et al. 2000; Lloyd et al. 2000),
the clustering of peak energies, the correlation between the
burst's peak energy and luminosity (Amati et al. 2000). It becomes
recognized that an additional thermal component may play a key
role and could solve these problems (e.g. Pe'er et al. 2006; Ryde
et al. 2006). It has also been pointed out  that a hybrid model
with both a thermal and non-thermal component can describe the
spectrum equally well as the Band function model (Band 1993), but
the former has a more physical meaning (Ryde 2005).  Recently, it
was suggested that a strong quasi-thermal component could result
from the internal dissipation occurring in the optically thick
inner parts of relativistic outflows (Rees \& \Meszaros 2005;
Pe'er, \Meszaros \& Rees 2006; Thompson et al. 2007).
Sub-photospheric shock dissipation can increase the radiative
efficiency of the outflow, significantly boosting the original
thermal photospheric component so that it may well dominate the
nonthermal component from optically-thin shocks occurring outside
the photosphere.

Neutrino emission from gamma-ray bursts has been predicted at
different stages of the relativistic outflow, such as the
precursor phase (e.g. Bahcall \& \Meszaros 2000; \Meszaros \&
Waxman 2001; Razzaque et al. 2003a,b; Razzaque et al. 2004; Ando
\& Beacom 2005; Horiuchi \& Ando 2008; Koers \& Wijers 2008), the
prompt emission phase (e.g. Waxman \& Bahcall 1997; Dermer \&
Atoyan 2003; Guetta et al. 2004; Murase \& Nagataki 2006; Gupta \&
Zhang 2007; Murase et al. 2006) and afterglow phase (e.g. Waxman
\& Bahcall 2001; Dai \& Lu 2001; Dermer 2002; Li et al. 2002;
Murase \& Nagataki 2006; Murase 2007; Dermer 2007). Based on the
broken power-law approximation for the spectrum of the prompt
emission, presumably from optically-thin internal shocks, a burst
of PeV neutrinos, produced by photomeson production, was predicted
to accompany the prompt gamma-ray emission if protons are present
and also accelerated in the shocks (Waxman \& Bahcall 1997). The
neutrino emission from proton-proton ($pp$) collisions was
generally thought to be negligible due to lower collision opacity
for optically-thin internal shocks. However, as we show below,  if
some part of the prompt emission arises from internal shocks
occurring in the optically-thick inner part of the outflow, as
indicated by the thermal emission, a lower energy ($\la10 {\rm
TeV}$) neutrino component may appear as a result of $pp$
collisions.

\section{The dissipative photosphere model}
The photosphere models have been widely discussed in relation to the
prompt emission of GRBs (e.g. Thompson 1994; Ghisellini \& Celotti
1999; Rees \& \Meszaros 2005; Thompson et al. 2007; Ioka et al.
2007). The potential advantage of photosphere models is that the
peak energy can be stabilized, which is identified as the thermal or
Comptonization thermal peak (see Ioka et al. 2007 and references
therein).  The photosphere radiation may also produce a large number
of electron-positron pairs, which may lead to a pair photosphere
beyond the baryon-related photosphere (e.g. Rees \& \Meszaros 2005),
and may also enhance the radiative efficiency (Ioka et al. 2007). On
the other hand, it is also suggested that the number of pairs
produced does not exceed the baryon related electrons by a factor
larger than a few (Pe'er et al. 2006). For simplicity,  we here only
consider the dissipation below the baryon-related photosphere, which
is more favorable for $pp$ neutrino production.

Following Rees \& \Meszaros (2005) and Pe'er et al. (2006), we
assume that during the early stage of the prompt emission,
internal shocks of the outflow occur at radii below the baryonic
photosphere. Initially, the internal energy is released at the
base of the outflow, $r_0\sim\alpha r_g=2\alpha GM/c^2$, where
$\alpha\ga1$ and $r_g$ is the Schwarzschild radius of a central
object of mass $M$. The internal energy is then converted to the
kinetic energy of the flow, whose bulk Lorentz factor grows as
$\gamma\sim r$ up to a saturation radius at $r_s\sim r_0\eta$,
where $\eta=L_0/(\dot{M}c^2)$ is the initial dimensionless
entropy, $L_0$ and $\dot{M}$ are the total energy and mass outflow
rates. Above the saturation radius, the observer-frame
photospherical luminosity decreases as
$L_\gamma(r)=L_0(r/r_s)^{-2/3}$ and the greater part of energy is
in kinetic form, $L_k\sim L_0$. If the dissipation is maintained
all the way to the photosphere, it will lead to an effective
luminosity $L_\gamma\sim\epsilon_d L_0$ and a temperature (Rees \&
\Meszaros 2005)
\begin{equation}
T_\gamma=\epsilon_d^{1/4}(r/r_s)^{-1/2}T_0=200 \epsilon_d^{1/4}
r_{11}^{-1/2} \Gamma_2^{1/2} L_{0,52}^{1/4}{\rm keV},
\end{equation}
where $\epsilon_d$ is the dissipation efficiency, $T_0$ is the
initial temperature of the fireball outflow and $\Gamma$ is the
bulk Lorentz factor of the outflow. The internal shock  occurs at
$R\simeq 2\Gamma r_s =6\times10^{10} \alpha\Gamma_2^2 (M/10
M_\odot) {\rm cm} $, for which the optical depth to Thomson
scattering by the baryon-related electrons is $\tau_{\rm
T}=\sigma_{\rm T} {L_k}/({4\pi R \Gamma^3 m_p c^3})=120 L_{k, 52}
R_{11}^{-1} \Gamma_2^{-3}$, where $\sigma_{\rm T}$ is the Thomson
scattering cross section and $L_k$ is the kinetic energy
luminosity.  The photosphere is further out, at radius
$R_{ph}=1.2\times10^{13} L_{k,52}\Gamma_2^{-3} {\rm cm}$. A
detailed calculation taking into account of the electron/positron
cooling and the Comptonization effect leads to a quasi-thermal
emission which peaks at energy $\sim300-500{\rm keV}$ for
dissipation at Thomson optical depth of $\tau_{\rm T}\sim10-100$
(Pe'er et al. 2006). This temperature is consistent with the
observed peak energies of prompt gamma-ray emission of a majority
of GRBs.

Assuming that a fraction of $\epsilon_B\simeq 0.1$ of the shock
internal energy is converted into magnetic fields, we have a
magnetic field $B'=2.5\times10^7
\epsilon_{B,-1}^{1/2}L_{k,52}^{1/2} R_{11}^{-1} \Gamma_2^{-1} {\rm
G}$. Protons  accelerated by internal shocks are assumed to have a
spectrum $dn/d\varepsilon_p\sim \varepsilon_p^{-p}$ with
$p\simeq2$, as often assumed for non-relativistic or
mildly-relativistic shock acceleration. The maximum proton energy
is set by comparing the acceleration time scale $t'_{acc}=\alpha
\varepsilon'_p/(e B' c)=4.4\times10^{-12} \alpha
(\frac{\varepsilon'_p}{1 {\rm GeV}})
\epsilon_{B,-1}^{-1/2}L_{k,52}^{-1/2} R_{11}\Gamma_2 \,{\rm s}$
with the energy-loss time scales. The synchrotron loss time is
$t'_{syn}=6\pi m_p^4 c^3/(\sigma_{\rm T}m_e^2 {\varepsilon'_p}
B'^2)=10^{-4} \epsilon_{B,-1}^{-1}L_{k,52}^{-1} R_{11}^2\Gamma_2^2
(\frac{\varepsilon'_p}{10^{8} {\rm GeV}})^{-1} \,{\rm s}$.
Assuming that the sub-photosphere emission at the dissipation site
peaks at $\varepsilon_\gamma=300{\rm keV}$ with a thermal-like
spectrum, the number density of photons in the comoving frame is $
n'_{\gamma}={L_\gamma}/({4\pi R^2 \Gamma^2 c
\varepsilon'_{\gamma}})=5\times10^{21} L_{\gamma,51} R_{11}^{-2}
\Gamma_2^{-1} ({\varepsilon_\gamma}/{300 {\rm keV}})^{-1} {\rm
cm^{-3}}$. The $p\gamma$ cooling time is approximately
$t'_{p\gamma}={1}/({\sigma_{p\gamma} n'_{\gamma} c
K_{p\gamma}})=10^{-4}L_{\gamma,51}^{-1} R_{11}^2
\Gamma_2(\frac{\varepsilon_\gamma}{300 {\rm keV}})\, {\rm s}$,
where $K_{p\gamma}\simeq 0.2$ is the inelasticity and
$\sigma_{p\gamma}=5\times10^{-28} {\rm cm^{2}}$ is the peak cross
section at the $\Delta$ resonance. By comparison with the
synchrotron loss time, it is found that the most effective cooling
mechanism for protons is the $p\gamma$ process for protons with
energies above the $p\gamma$ threshold, but below $\sim10^8 {\rm
GeV}$. Equating $t'_{acc}=t'_{p\gamma}$, we obtain the maximum
proton energy in the shock comoving frame
\begin{equation}
\varepsilon'_{p, max}=10^7 \alpha_1^{-1}
\epsilon_{B,-1}^{1/2}L_{k,52}^{1/2}L_{\gamma,51}^{-1}R_{11}(\frac{\varepsilon_\gamma}{300
{\rm keV}}) \, {\rm GeV} .
\end{equation}

\section{Proton and meson cooling}
The shock-accelerated protons produce mesons via $pp$ and
$p\gamma$ interactions. Since the meson multiplicity in $pp$
interactions is about  1 for pions while 0.1 for kaons, neutrinos
contributed by pion decay are dominant when the cooling effect of
pions is not important, which is applicable to the low energy pp
neutrinos. Therefore we here consider only pion production in $pp$
interactions. The pion production by $pp$ interaction in the
sub-photosphere dissipation is efficient since the cooling time in
the shock comoving frame,
\begin{equation}
t'_{pp}=1/(\sigma_{pp} n'_p c K_{pp})=0.008
L_{k,52}^{-1}R_{11}^2\Gamma_2^2 \, {\rm s},
\end{equation}
can be shorter than the shock  dynamic time, $t'_{dyn}=R/\Gamma
c=0.03 R_{11}\Gamma_2^{-1} {\rm sec}$, where
$\sigma_{pp}=4\times10^{-26} {\rm cm^2}$ is the cross section for
$pp$ interactions, $n'_p={L_k}/({4\pi R^2 \Gamma^2 m_p
c^3})=2\times10^{17} L_{k,52}R_{11}^{-2}\Gamma_2^{-2} {\rm
cm^{-3}}$ is the proton number density, and $K_{pp}\simeq0.5$ is
the inelasticity. Protons also cool through Bethe-Heitler
interactions ($p\gamma\rightarrow p e^+ e^-$) and $p\gamma$
interactions when the target photon energy seen by the protons is
above the threshold energy for each interaction. Denoting by
$n(\epsilon_\gamma)d\epsilon_\gamma$ the number density of photons
in the energy range $\epsilon_\gamma$ to
$\epsilon_\gamma+d\epsilon_\gamma$, the cooling time in the shock
comoving frame for $p\gamma$ and Bethe-Heitler cooling  processes
are given by
\begin{equation}
t'_{\{p\gamma, BH\}}=\frac{c}{2\Gamma_p^2}
\int_{\epsilon_{th}}^\infty d\epsilon \sigma (\epsilon)
K(\epsilon)\epsilon\int_{\epsilon/2\Gamma_p}^{\infty} dx x^{-2}
n(x),
\end{equation}
where $\Gamma_p=\varepsilon'_p/m_p c^2$, $\sigma$ and $K$ are
respectively the cross section and the inelasticity for $p\gamma$
(or Bethe-Heitler) process. As a rough estimate, the Bethe-Heitler
cooling time is
\begin{equation}
\frac{t'_{BH}}{t'_{pp}}=0.5 (\frac{(28/9) {\rm
ln}40-218/27}{(28/9) {\rm ln}2k-218/27
})(\frac{L_{k,52}}{L_{\gamma, 51}})(\frac{\varepsilon'_\gamma}{3
{\rm keV}})
\end{equation}
when $k\equiv\varepsilon'_p \varepsilon'_\gamma/(m_p m_e c^4)$ is a
large value (a good approximation when $k\ga10$, Chodorowski et al.
1992), where $\varepsilon'_\gamma$ is the thermal peak energy of
photons in the comoving frame. So when the proton energy is larger
than $\varepsilon'^{(1)}_{p,b}=1500 (\frac{\varepsilon'_\gamma}{3
{\rm keV}})^{-1}{\rm GeV}$, the Bethe-Heitler cooling dominates over
the $pp$ cooling. At even higher energies near the threshold for
$p\gamma$ interactions at $\varepsilon'^{(2)}_{p,b}=6\times10^4
(\frac{\varepsilon'_\gamma}{3 {\rm keV}})^{-1}{\rm GeV}$, $p\gamma$
cooling becomes increasingly dominant.

\begin{figure}
\centering \epsfig{figure=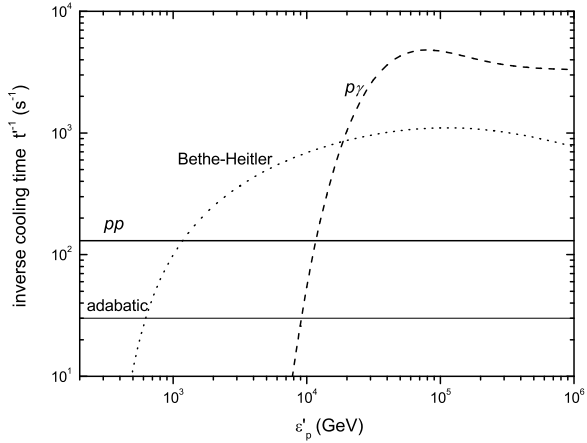,width=9cm} \caption{Inverse of
proton cooling time scales in the comoving frame of the internal
shock as functions of proton energy. The straight solid line,
dotted line and dashed line are for proton-proton collision,
Bethe-Heitler cooling and $p\gamma$ processes respectively. Also
shown is the  cooling time scale due to adiabatic expansion (which
is equal to the dynamic timescale $t'_{dyn}$ given in the text).
The parameters used in the plot are $L_k=10^{52} {\rm erg
s^{-1}}$, $\Gamma=100$, $R=10^{11}{\rm cm}$ and
$\varepsilon_\gamma=300{\rm keV}$.}
\end{figure}

We compare these  three cooling time scales for protons in Fig.1 for
representative  parameters, using more accurate cross sections for
$p\gamma$ and Bethe-Heitler processes in Eq.(4). For photopion
production cross section, we take the Lorentzian form for the
resonance peak (M\"ucke et al. 2000) plus a component contributed by
multi-pion production at higher energies, while for Bethe-Heitler
process we use the cross section given by Chodorowski et al. (1992).
The number density of photons  used in the calculation has been
assumed to have a blackbody distribution. The numerical result
confirms that that $pp$ cooling and $p\gamma$ cooling are dominant,
respectively, at lowest and highest energies, while at the
intermediate energies, the Bethe-Heitler cooling  is the dominant
one.

Defining the total cooling time for protons as
$t'_p=1/({t'}^{-1}_{pp}+{t'}^{-1}_{BH}+{t'}^{-1}_{p\gamma})$, the
total energy loss fraction of protons is $\eta_p={\rm
Min}\{t'_{dyn}/t'_p, 1\}$. The fractions of energy loss by $pp$
and $p\gamma$ processes are respectively,
\begin{equation}
\left \{
\begin{array}{ll}
\zeta_{pp}={t'}^{-1}_{pp}/({t'}^{-1}_{pp}+{t'}^{-1}_{BH}+{t'}^{-1}_{p\gamma})\\
\zeta_{p\gamma}={t'}^{-1}_{p\gamma}/({t'}^{-1}_{pp}+{t'}^{-1}_{BH}+{t'}^{-1}_{p\gamma}).
\end{array} \right.
\end{equation}

The cooling of secondary pions may also affect the neutrino
production efficiency if they suffer from cooling before decaying
to secondary products. The pions suffer from radiative cooling due
to both synchrotron emission and inverse-Compton emission. The
total radiative cooling time is $t'_{\pi, rad}={3 m_\pi^4
c^3}/[{4\sigma_T m_e^2 \epsilon'_{\pi} U'_B(1+f_{\rm
IC})}]\simeq0.002 ({\epsilon'_{\pi}}/{1 {\rm TeV}})^{-1}
\epsilon_{B,-1}^{-1}L_{k,52}^{-1} R_{11}^2 \Gamma_2^2 \,{\rm s}$,
where $U'_B$ is the energy density of the magnetic filed in the
shock region and $f_{\rm IC}\la1$ is the correction factor
accounting for the inverse-Compton loss. The pions also cool due
to collisions with protons (Ando \& Beacom 2005). The cooling for
this hadronic process is $t'_{\pi,
had}={\epsilon'_{\pi}}/({c\sigma_{\pi p} n'_p \Delta
\epsilon'_{\pi}})=0.006 L_{k,52}^{-1} R_{11}^2 \Gamma_2^2 \, {\rm
s}$, where $\sigma_{\pi p}=5\times10^{-26}{\rm cm^2}$ is the cross
section for meson-proton collisions and $\Delta
\epsilon'_{\pi}=0.5\epsilon'_{\pi}$ is the energy lost by the
meson per collision. The suppression of neutrino emission due to
cooling of pions can be obtained by comparing the  cooling time
$t'_{\pi, rad}$ or $t'_{\pi, had}$ with the lifetime of pions
$\tau'_{\pi}=\gamma_{\pi} \tau=1.9\times10^{-4} (\epsilon'_{\pi}/1
{\rm TeV}) \,{\rm s}$ in the shock comoving frame, where
$\gamma_{\pi}$ and $\tau$ are the pion Lorentz factor and proper
lifetime. This defines two critical energies for pions, above
which the effect of radiative cooling or hadronic cooling starts
to suppress the neutrino flux, i.e. $\epsilon'_{\pi, rad}=3
\epsilon_B^{-1/2}L_{k,52}^{-1/2}R_{11}\Gamma_2 {\rm TeV}$ and
$\epsilon'_{\pi, had}=30 L_{k,52}^{-1}R_{11}^2\Gamma_2^2 {\rm
TeV}$.  The total cooling time of pions is $t'_{\pi,
c}=1/({t'^{-1}_{\pi, rad}}+{t'^{-1}_{\pi, had}})$ and the total
suppression factor on the neutrino flux due to pion cooling is
(Razzaque et al. 2004)
\begin{equation}
\zeta_{\pi}={\rm Min}\{t'_{\pi, c}/\tau'_{\pi}, 1\}.
\end{equation}

\begin{figure}
\centering \epsfig{figure=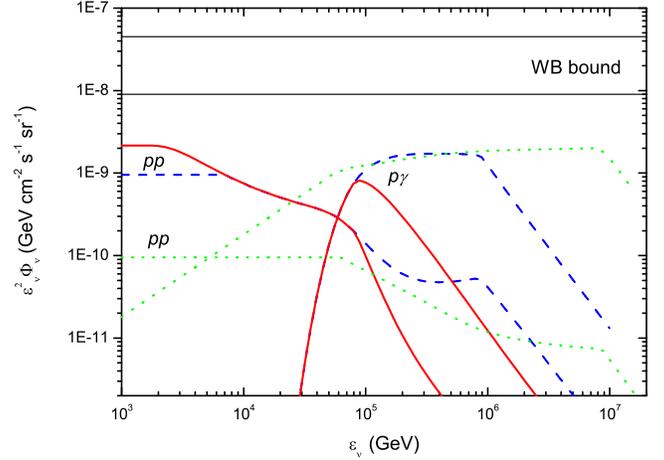,width=10cm} \caption{Diffuse muon
neutrino flux on Earth contributed by $pp$ and $p\gamma$
interactions from the dissipative photosphere of GRBs, assuming
that energy in accelerated protons in one burst is
$E_p=1.5\times10^{53}$ erg during the dissipative photosphere
phase, the bulk Lorentz factor $\Gamma=100$, the GRB rate $R_{\rm
GRB}(0)=1 {\rm Gpc^{-3} yr^{-1}}$ and $f_z=3$. The solid (red) and
dashed (blue)  lines are for shock dissipation at $R=10^{11} {\rm
cm}$ and $10^{12} {\rm cm}$ respectively. The dotted (green) line
is for shock at photosphere radius $10^{13} {\rm cm}$ with a
broken power-law photon spectrum assumed (see the text for
details). The solid lines denote the $pp$ neutrino component,
while the dashed lines denote the $p\gamma$ neutrino component.
Also shown is the Waxman-Bahcall (WB) bound (Waxman \& Bahcall
1999).}
\end{figure}
\section{Neutrino flux from GRBs}
The total energy emitted in
neutrinos from $pp$ or $p\gamma$ processes per GRB is,
respectively,
\begin{equation}
{\varepsilon_\nu^2}J_{\{pp,p\gamma\}}(\varepsilon_\nu)=\frac{1}{8}
\frac{E_p \eta_{p}(\varepsilon_\nu)
{\zeta_{\{pp,p\gamma\}}(\varepsilon_\nu)\zeta_{\pi}(\varepsilon_\nu)}}{{\rm
ln}(\varepsilon'_{p,max}/\varepsilon'_{p,min})},
\end{equation}
where $E_p$ is the energy in accelerated protons in one burst
during the dissipative photosphere phase, $\varepsilon'_{p,max}$
and $\varepsilon'_{p,min}$ are the maximum and minimum energies of
acceleration protons. In the absence of pion cooling loss, the
neutrinos produced by pion decay carry $1/8$ of the energy lost by
protons to pion production, since charged  and neutral pions are
produced with roughly equal probability and muon neutrinos carry
roughly $1/4$ of the pion energy in pion decay \footnote{As an
approximate estimate, we have neglected the effect of multi-pion
production, the muon contribution decay to the neutrino flux and
the neutrino oscillation effect, which may affect the estimate of
the $pp$ neutrino flux within a factor of 2 (the factor, however,
could be larger for $p\gamma$ neutrino flux).}. The mean pion
energy is about $20\%$ of the energy of the proton producing the
pion, so the mean energy of neutrinos is $\varepsilon_\nu\simeq
0.05 \varepsilon_p$. Assuming that protons are efficiently
accelerated in shocks  with an energy density of $U'_p=10
U'_\gamma$, the number of TeV neutrinos from one GRB is about
$N_\nu=0.1 (\Phi_\gamma/10^{-4} {\rm erg cm^{-2}})$ for
$\eta_p\simeq1$, according to Eq.(8). So only from very strong
bursts with gamma-ray fluence $\Phi_\gamma\ga 10^{-3} {\rm erg
cm^{-2}}$, which are very rare events, can neutrinos from single
GRB be detected.

The aggregated muon neutrino flux from all GRBs is approximately
given by
\begin{equation}
\begin{array}{ll}
{\varepsilon_\nu^2}\Phi_{\{pp,p\gamma\}}(\varepsilon_\nu)\simeq(\frac{c}{4\pi
H_0}) {\varepsilon_\nu^2}J_{\{pp,p\gamma\}}(\varepsilon_\nu)
R_{\rm
GRB}(0) f_z \\
=1.5\times10^{-9} E_{p, 53} (\frac{R_{\rm GRB}(0)}{1 {\rm Gpc^{-3}
yr^{-1}}})(\frac{f_z}{3}) \eta_{p}
{\zeta_{\{pp,p\gamma\}}}\zeta_{\pi} {\rm GeV cm^{-2} s^{-1}
sr^{-1}},
\end{array}
\end{equation}
where $f_z$ is the correction factor for the contribution from
high redshift sources and $R_{\rm GRB}(0)$ is the overall GRB rate
at redshift $z=0$. Assuming that GRB rate traces the
star-formation rate in the Universe, the calculation gives
$f_z\simeq3$ (Waxman \& Bahcall 1999). It is not clear how
efficiently the protons are accelerated in GRB shocks. Assuming an
optimistic case that protons are efficiently accelerated in shocks
and that half of the kinetic energy dissipation occurs below the
photosphere\footnote{This is based on the analysis by Ryde (2006)
and also in a very recent paper by Ryde \& Pe'er (2008,
arXiv:0811.4135v1), who find that the thermal photons carry a
fraction of ~30\% to more than 50\% of the prompt emission
energy.}, we take a mean value $E_p = 1.5\times10^{53} {\rm ergs}$
for the isotropic equivalent energy in accelerated protons in one
GRB during the dissipative photosphere phase, based on a typically
used value $L_k=10^{52} {\rm erg s^{-1}}$ for the isotropic
kinetic energy luminosity and a typical long GRB duration of
$\Delta T=30$ s. The GRB rate\footnote{There is large uncertainty
in the estimate of the local GRB rate. Some people suggest a lower
GRB rate based on the analysis of {\it Swift} bursts with  $R_{\rm
GRB}(0)=0.05-0.27 {\rm Gpc^{-3} yr^{-1}}$ (e.g. Guetta \& Piran
2007; Le \& Dermer 2007) , while others get a higher rate
comparable to earlier estimate before {\it Swift} (e.g. Liang et
al. 2007).}  at redshift $z=0$ is taken to $R_{\rm GRB}(0)=1 {\rm
Gpc^{-3} yr^{-1}}$ (Guetta et al. 2005; Liang et al. 2007). The
isotropic luminosity is taken to be $L_\gamma=10^{51}{\rm erg}$
\footnote{Some observations have indicated rather high radiative
efficiency, and the importance of $pp$ neutrinos may be reduced if
$L_k$ is smaller than $10L_\gamma$.}. The energy-dependent
neutrino flux contributed by $pp$ and $p\gamma$ interactions are
plotted in Fig.2 for a set of representative parameters of the
dissipative photosphere model and three different dissipation
radii. If the kinetic energy is dissipated at  radius of
$R=10^{11}{\rm cm}$, the calculation (the red solid curves) shows
that at energies below tens of TeV, the neutrino flux is dominated
by a $pp$ component. Taking the detection probability of
$P_{\nu\mu}=10^{-6}(\varepsilon_\nu/1 {\rm TeV})$ for TeV
neutrinos (Gaisser et al. 1995), the expected flux of upward
moving muons contributed by this $pp$ component is about 8-10
events each year for a ${\rm km^3}$ neutrino detector, such as
Icecube. {We can also estimate the atmospheric neutrino background
expectation in coincidence with these GRB sources, noting that the
search for neutrinos accompanying GRBs requires that the neutrinos
are coincident in both direction and time with gamma-rays. Taking
an average GRB duration of $\simeq 30 $ s, an angular resolution
of Icecube of $\simeq1^\circ$ and the atmospheric neutrino
background flux of $\simeq 10^{-4} {\rm GeV cm^{-2} s^{-1}
sr^{-1}}$ at 1 TeV (Ahrens et al. 2004), the atmospheric neutrino
background expectation is $\simeq5\times10^{-3}$ events from 500
GRBs (in one year). Such a low background  in coincident with GRBs
allows the claim of detection of TeV neutrinos from GRB sources.}
At energies from a few TeV to tens of TeV, the Bethe-Heilter
cooling suppresses the $pp$ cooling, resulting in a steepening at
several TeV in the neutrino spectrum. At energies above the
threshold for $p\gamma$ interactions, the neutrino from $p\gamma$
process is heavily suppressed due to the strong radiative cooling
of secondary pions. For a larger dissipation radius at $R=10^{12}
{\rm cm}$ (the blue dashed curves), the neutrino emission flux
from $p\gamma$ process is no longer suppressed and in this case
both $pp$ and $p\gamma$ neutrino components could be detected by
${\rm km^3}$ detectors. { We also calculated the neutrino flux,
shown by the green dotted lines in Fig. 2, for shock at
photosphere radius $R_{ph}$, assuming that the radiation spectrum
is a broken power-law (due to Comptonization) peaking at
$\varepsilon_\gamma=100{\rm keV}$, with lower energy and higher
energy photon indexes given by $-1$ and $-2$ respectively.  In
this case, the $pp$ neutrino flux becomes small (may be marginally
detectable), while the $p\gamma$ neutrino flux spectrum is similar
to the analytic result obtained by Waxman \& Bahcall (1997), as
expected for a broken power-law photon spectrum.  Note that in one
burst the shock dissipation could be continuous from small to
large radii, as indicated by the larger variability timescales
seen in GRBs. By comparing the three cases of different
dissipation radii in Fig.2, one can see that as the shock radius
increases, the neutrino emission from $pp$ component decreases,
while the $p\gamma$ component increases until it reaches the
saturation level. The total $pp$ neutrino flux from such
continuous dissipation is thus contributed predominantly by the
deepest internal shocks below the photosphere. In the whole
neutrino spectrum, a ``valley" is seen between the  $pp$ and
$p\gamma$ components of the spectrum, which may be a potential
distinguish feature of the sub-photosphere dissipation effect.}

\section{Discussions and Conclusions}
Waxman \& Bahcall (1997) as well as later works have studied the
neutrino emission from the photomeson process during the prompt
internal shocks of GRBs, assuming that the radiation in the shock
region has a broken power-law nonthermal spectrum. It was found
that the neutrino emission peaks at energies above $100$ TeV.
Towards lower energies, the neutrino emission intensity decreases
as $\varepsilon_{\nu}^2 \Phi_{\nu}\sim \varepsilon_\nu$ (Waxman \&
Bahcall 1997). However, if internal shocks, especially at the
early stage of the prompt emission, occur below the photosphere, a
quasi-thermal spectrum will arise. In this {\it Letter}, we have
discussed the neutrino emission associated with the dissipative
photosphere that produces such prompt thermal emission. We find
that $pp$ interaction process becomes important for
shock-accelerated protons and provides a new neutrino component,
which dominates at energies below tens of TeV. The neutrino
emission from photopion process of protons interacting with the
sub-photosphere radiation could be significantly suppressed due to
radiative cooling of secondary pions, when the dissipation radius
is relatively small. Nevertheless, the total contribution by
photopion process will not be suppressed since the shock
dissipation could be continuous and occur at large radii as well.
Although TeV neutrinos may also be produced during the early
precursor stage of GRB, i.e. before the jet breaking out the
progenitor star (e.g. Razzaque et al. 2004; Ando \& Beacom 2005),
we want to point out that the TeV neutrino component discussed
here can be distinguished from them, because in our case the
neutrino emission is associated in time with the prompt emission.

{\acknowledgments After this work  has been completed and later
put onto the arXiv website (arXiv:0807.0290),  we became aware
that K. Murase was also working on the sub-photosphere neutrino
independently (K. Murase, 2008, arXiv:0807.0919). XYW would like
to thank P. \Meszaros, S. Razzaque, K. Murase, E. Waxman, Z. Li
and K. Ioka for useful comments or discussions. This work is
supported by the National Natural Science Foundation of China
under grants 10221001, 10403002 and 10873009, the National Basic
Research Program of China (973 program) under grants No.
2007CB815404 and 2009CB824800.}

\end{document}